\documentclass[twocolumn,showpacs,preprintnumbers,amsmath,amssymb,prl]{revtex4-1}
\usepackage{amssymb}
\usepackage{epsfig,amsmath,graphics}
\usepackage{epstopdf}
\usepackage{graphicx}

\renewcommand{\v}[1]{\ensuremath{\mathbf{#1}}}

\newcommand{\ket}[1]{\ensuremath{\left|#1\right>}}

\def\be{\begin{equation}}
\def\ee{\end{equation}}

\def\keff{k_{\text{eff}}}

\def\nano{\text{n}}
\def\micro{\mu}
\def\milli{\text{m}}
\def\centi{\text{c}}
\def\kilo{\text{k}}
\def\Mega{\text{M}}
\def\Giga{\text{G}}

\def\Kelvin{\text{K}}
\def\radians{\text{rad}}
\def\seconds{\text{s}}
\def\Hz{\text{Hz}}
\def\meters{\text{m}}

\def\Watts{\text{W}}
\def\Gauss{\text{G}}

\def\phiEarth{\phi_E}
\def\OmegaEarth{\Omega_E}
\def\OmegaRCS{\Omega_\text{C}}
\def\phiRCS{\phi_\text{C}}

\begin{document}


\title{Multi-axis inertial sensing with long-time point source atom interferometry}

\author{Susannah M. Dickerson}
\affiliation{Department of Physics, Stanford University, Stanford, California 94305}
\author{Jason M. Hogan}
\affiliation{Department of Physics, Stanford University, Stanford, California 94305}
\author{Alex Sugarbaker}
\affiliation{Department of Physics, Stanford University, Stanford, California 94305}
\author{David M. S. Johnson}
\affiliation{Department of Physics, Stanford University, Stanford, California 94305}
\author{Mark A. Kasevich}
\affiliation{Department of Physics, Stanford University, Stanford, California 94305}

\date{\today}

\begin{abstract}
We show that light-pulse atom interferometry with atomic point sources and spatially resolved detection enables multi-axis (two rotation, one acceleration) precision inertial sensing at long interrogation times. Using this method, we demonstrate a light-pulse atom interferometer for $^{87}$Rb with $1.4~\text{cm}$ peak wavepacket separation and a duration of $2\,T = 2.3~\seconds$. The inferred acceleration sensitivity of each shot is $6.7 \times 10^{-12} g$, which improves on previous limits by more than two orders of magnitude. We also measure the Earth's rotation rate with a precision of 200 nrad/s.
\end{abstract}


\maketitle

Light-pulse atom interferometry enables precision tests of gravity \cite{Fixler2007, Dimopoulos2008b, Hogan2009} and electrodynamics \cite{Bouchendira2011} as well as practical applications in inertial navigation, geodesy, and timekeeping. Phase shifts for light-pulse atom interferometers demonstrate sensitivity to the initial velocity distribution of the atom source, often resulting in inhomogeneous dephasing that washes out fringe contrast \cite{Gustavson1997}. In this Letter, we show that use of spatially resolved imaging in combination with an initially spatially localized atomic source allows direct characterization of these phase shifts.  We refer to this technique as point source interferometry (PSI).

The contrast loss associated with such inhomogeneous dephasing is not fundamental, but is a consequence of atom detection protocols that average over velocity-dependent phase shifts. With PSI we establish a correlation between velocity and position and use spatially-resolved detection to form an image of the ensemble that reveals its velocity-dependent phase structure.  A simple way to realize this correlation is through ballistic expansion of the ensemble.  In the limit that the ensemble size at detection is much larger than its initial size, each atom's position is approximately proportional to its initial velocity. Consequently, any initial velocity-dependent phase shift results in a spatial variation of the interferometer phase, yielding a position-dependent population difference between the two output ports of the interferometer.

An important example of velocity sensitivity is due to rotation of the interferometer laser beams \cite{Peters2001, Hogan2009}.  Rotation at a rate $\v{\Omega}$ leads to a phase shift (Table \ref{Tab:phases}, term 2) that depends on $\left(v_x,v_y\right)$, the initial transverse velocity of the atom.  In a rotating frame, this effect may be interpreted as a Coriolis acceleration. PSI also allows observation of longitudinal velocity-dependent phase shifts in asymmetric atom interferometers \cite{Muntinga2013} (e.g., Table \ref{Tab:phases}, term 3).

\begin{table}
	\begin{tabular}{c c c}
 		Term & Phase Shift & Size ($\radians$)\\
		\hline
		1 & $  \keff g \, T^2$ & $2.1 \times 10^8$ \\
        2 & $ 2\v{\keff} \cdot \left(\v{\Omega} \times \v{v}\right) T^2$ & $5.1$ \\
        3 & $  \keff v_z \delta T$ & $3.5$ \\
        4 & $ \tfrac{\hbar\keff^2}{2 m} T_{zz} T^3$ & $0.44$ \\
        5 & $  \keff T_{zi} \left(x_i + v_{i} T \right) T^2$ & $0.18$\\
        6 & $  \tfrac{1}{2}\keff \alpha \left(v_x^2 + v_y^2\right) T^2$ & $0.04$ \\
	\end{tabular}
\caption{Velocity-dependent phase shifts and their sizes assuming the following: $\keff = 2 k = 2 \cdot 2\pi/780~\nano \meters$, $T = 1.15~\seconds$, initial velocity spread $v_i = 2~\milli \meters/\seconds$ ($50~\nano\Kelvin$), initial positions $x_i = 200~\micro \meters$, $\left|\v{\Omega}\right|=60~\micro \radians/\seconds$, gravity gradient tensor components $T_{zi}=3075~\text{E}$, interferometer pulse timing asymmetry $\delta T = 100~\micro\seconds$, and wavefront curvature $\alpha = (\lambda/10) / \centi \meters^2$. Note that for $T_{zx}, T_{zy} = 50~\text{E}$ the size of term 5 is significantly smaller. The acceleration (term 1) and gravity curvature (term 4) phase shifts are shown for reference. \label{Tab:phases}}
\end{table}

\begin{figure}
\centering
\includegraphics[width=\columnwidth]{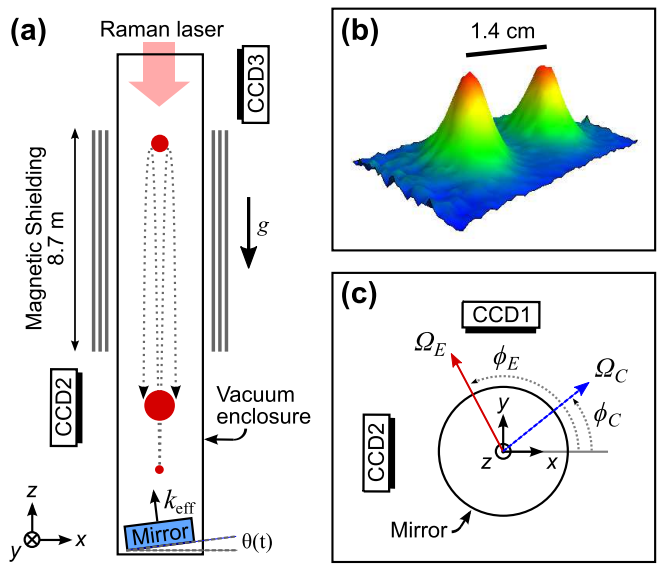}
\caption{(a) Schematic diagram of the apparatus, viewed from the side.  The atom cloud (red circle) is cooled and launched from below the magnetically-shielded interferometry region.   The two interferometer output ports are imaged by both perpendicular cameras (CCD1 and CCD2).  All interferometry pulses are delivered from the top of the tower and are retroreflected off a mirror (at angle $\theta(t)$) resting on a piezo-actuated tip-tilt stage. (b) Image of the ensemble after a beamsplitter pulse showing the separation between two halves of the atomic wavepacket. For this shot we launched the atoms with extra velocity to reach CCD3. (c) Top view of the tip-tilt stage and lower cameras with the direction and magnitude of the Earth rotation $\v{\OmegaEarth}$ and an (arbitrary) applied counter-rotation $\v{\OmegaRCS}$.}
\label{Fig:Apparatus}
\end{figure}

To demonstrate PSI, we induce a velocity-dependent phase shift in a $^{87}$Rb Raman light-pulse atom interferometer. We launch cold atoms from the bottom of a 10-meter tall vacuum enclosure (Fig.~\ref{Fig:Apparatus}a) and apply a three-pulse accelerometer sequence ($\pi/2-\pi-\pi/2$) \cite{Kasevich1991}. The first pulse serves as an atom beamsplitter, coherently driving the atoms into a superposition of states $\ket{F=1; p}$ and $\ket{F = 2; p+\hbar \keff}$ with momentum difference $\hbar \keff = 2 \hbar k$. Over the subsequent $T = 1.15~\seconds$ interrogation interval, the two parts of the atom's wave function separate vertically by $\tfrac{\hbar\keff}{m}T = 1.4~\centi \meters$ (Fig.~\ref{Fig:Apparatus}b), at which time a mirror pulse reverses the relative momenta and internal states. After an identical drift time, a final beamsplitter pulse interferes the atom wave packets.  We then image the atom fluorescence using a pair of CCD cameras located below the interferometry region (Fig.~\ref{Fig:Apparatus}c). By the time of imaging, $2.6~\seconds$ after launch, the $50~\nano \Kelvin$ atomic source has expanded to $30$ times its original size, establishing the position-velocity correlation necessary for PSI.

\begin{figure*}
\centering
\includegraphics[width=\textwidth]{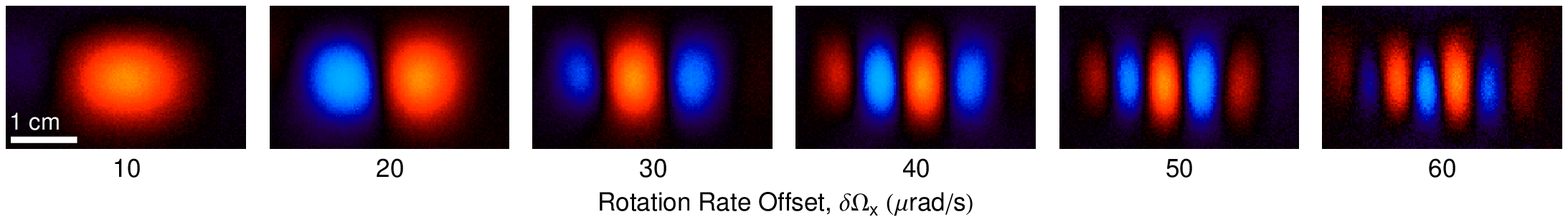}
\caption{Spatial fringes on the atom population observed on CCD2 versus rotation rate offset $\delta\Omega_x$. Blue versus red regions show anti-correlation in atom population. The second output port, with fringes $\pi~\text{rad}$ out of phase, is not shown. Each image is the second-highest variance principle component arising from a set of 20 measurements \cite{supplementalMaterial}.}
\label{Fig:FringeImages}
\end{figure*}

We imprint a velocity-dependent phase shift by rotating the atom interferometer laser beam axis at a tunable rate $\v{\delta\Omega}$.  Figure~\ref{Fig:FringeImages} shows typical detected atom distributions for several different values of $\delta\Omega_x$.

\begin{figure}
\centering
\includegraphics[width=\columnwidth]{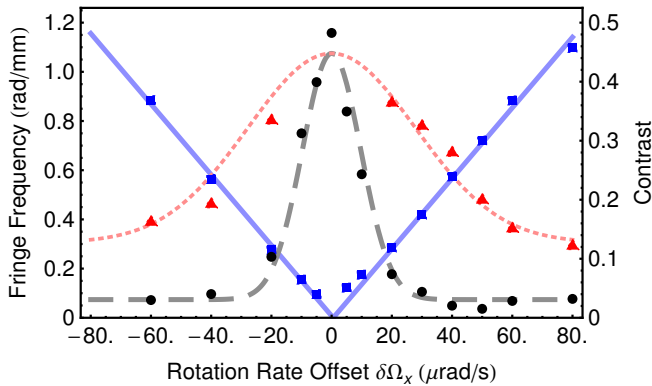}
\caption{Fringe spatial frequency (blue squares, solid line) and contrast versus applied rotation for the data in Fig. \ref{Fig:FringeImages}. The fitted slope of the fringe spatial frequency is consistent with term 2 of Table~\ref{Tab:phases} to $<10\%$.  Fringe contrast is observed over a wide range of rotation rates (red triangles, dotted line Gaussian fit), while the contrast from integration detection decays rapidly (black circles, dashed line Gaussian fit).}
\label{Fig:Contrast}
\end{figure}

The velocity-dependent phase gradient we observe in Fig.~\ref{Fig:FringeImages} is proportional to the applied rotation rate (Fig.~\ref{Fig:Contrast}).  For faster rates, the phase shift is large enough that multiple fringe periods appear across the ensemble.  Without spatially resolved detection, averaging over these fringes would yield negligible contrast.  With PSI, we realize record duration atom interferometry, even in the presence of large rotation rates.

To create the cold atomic source, we load $4 \times10^9$ atoms from a magneto-optical trap into a plugged quadrupole trap, where we evaporate with a microwave knife \cite{Davis1995, Dubessy2012}. A magnetic lensing sequence in a time-orbiting potential (TOP) trap collimates the atom source in 3D, cooling and expanding the cloud while maintaining high phase space density \footnote{The procedure is similar in principle to $\delta$-kick cooling \cite{Ammann1997}, but uses the atoms' continuous expansion over $\sim 100 \, \milli \seconds$ against a shallow ($\sim 5 \, \Hz$) harmonic trap \cite{Monroe1990} rather than a short (few ms) impulse \cite{Muntinga2013}. The magnetic fields are rapidly turned off when the atoms have reached their minimum velocity (maximum expansion) in all three dimensions}.  The final cloud contains $4 \times 10^6$ atoms at $50~\nano \Kelvin$ with an initial radius of $200 \, \micro \meters$.  Alternatively, we can produce clouds at $3 \, \nano \Kelvin$ with $10^5$ atoms and an initial radius of $30 \, \micro \meters$ by evaporating in a TOP trap with a microwave knife prior to the magnetic lensing sequence.

A microwave pulse transfers the ultracold atoms into a magnetically-insensitive Zeeman sublevel.  They are then coherently launched with an optical lattice \cite{Denschlag2002}, which transfers $2386$ photon momenta with a peak acceleration of $75 \, \text{g}$. They enter the interferometer region, a $10\, \centi \meters$ diameter, $8.7 \, \meters$ long aluminum vacuum tube. A solenoid wound around the tube provides a bias magnetic field, and three layers of magnetic shielding suppress the environmental field to $< 1\, \milli \Gauss$ \cite{Dickerson2012}.

A small fraction of the atoms are launched into $\pm 2 \hbar k$ momentum states. We purify the ensemble's vertical momentum with a $135 \, \micro \seconds$ Raman $\pi$-pulse, which transfers a $25 \, \nano \Kelvin$ ($0.1 \, \hbar k$) subset of the ensemble into $\ket{F=1}$. A short pulse resonant with $\ket{F=2} \rightarrow \ket{F'=3}$ blows away atoms that did not transfer.

A pair of fiber-coupled $1\, \Watts$ tapered amplifiers (TAs) generate the retroreflected interferometer pulses. The seeds for the two TAs are derived from a common source cavity-stabilized to a linewidth of $< 1\, \text{kHz}$ and detuned $1.0 \, \Giga \Hz$ blue from the $780 \,\nano \meters$ $\text{D}_2$ line ($\ket{F=2} \rightarrow \ket{F'=3}$). The seed for one TA passes through a fiber phase modulator that generates the $6.8 \, \Giga \Hz$ sideband necessary for Raman interferometry. An acousto-optic modulator (AOM) chirps the other seed to correct for the atoms' Doppler shift. The output of the TAs are combined on a polarizing beamsplitter cube, and the copropagating beams are diffracted by an AOM that acts as a fast optical switch. The beamsplitter and mirror pulses are $35 \, \micro \seconds$ and $70 \, \micro \seconds$ in duration, respectively. The beams have a $2 \, \centi \meters$ $1/e^{2}$ intensity radial waist. The relative power of the two beams is chosen empirically to suppress intensity-dependent detunings by balancing AC Stark shifts (to $< 2 \, \kilo \Hz$).

Prior to detection, we spatially separate the output ports by applying a short pulse ($\sim\!50$ photon recoils) resonant with $\ket{F=2} \rightarrow \ket{F'=3}$. We wait $50 \, \milli \seconds$ before simultaneously halting and imaging the atoms with a $2\, \Mega \Hz$ red-detuned beam. The atoms are nearly at rest after the first $300\, \micro \seconds$ of the $5 \, \milli \seconds$ imaging time. The scattered light is collected by two orthogonal CCD cameras, each with a numerical aperture of $0.25$ (Fig.~\ref{Fig:Apparatus}c). The time from initial atom loading to the final image is $20 \, \seconds$.

We precisely control the direction of the interferometer beams with an in-vacuum, piezo-actuated tip-tilt stage onto which the retroreflection mirror is kinematically constrained. The stage has $1 \, \nano\radians$ measured precision and a range of $400 \, \micro \radians$. The stage platform is secured kinematically to three nanopositioners (Nano-OP30; Mad City Labs) by stiff springs. The nanopositioners are bolted to the vacuum enclosure, which is anchored to the vibrationally-quiet ($10^{-8} \, g/\sqrt{\Hz}$) concrete floor.

The rotation of the Earth is a significant source of velocity-dependent phase shifts. At our latitude in Stanford, California, the effective rate is $\OmegaEarth = 57.9 \, \micro \radians /\seconds$, which induces fringes of periodicity similar to the highest rotation rate in Fig.~\ref{Fig:FringeImages}. With the tip-tilt stage we apply a compensating rotation of equal and opposite magnitude ($\v{\OmegaRCS} = -\v{\OmegaEarth}$) to eliminate these phase shifts \cite{Gustavson1997, Hogan2009, Lan2012}.  We implement this rotation by incrementing the mirror's angle in discrete steps between each interferometer pulse. In Figs.~\ref{Fig:FringeImages} and \ref{Fig:Contrast} we add a variable rotation rate $\delta\Omega_x$ to this nominal rotation compensation vector.

\begin{figure}
\centering
\includegraphics[width=\columnwidth]{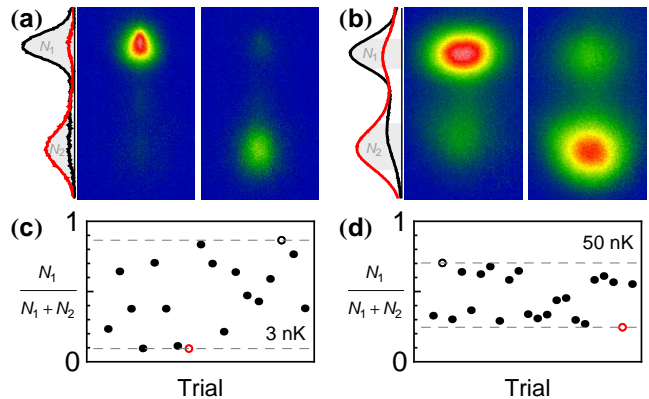}
\caption{Images of the interferometer output ports using (a) $3~\nano \Kelvin$ and (b) $50~\nano\Kelvin$ atom sources with rotation compensation ($\v{\OmegaRCS} = -\v{\OmegaEarth}$). The upper (lower) port consists of $N_1$ ($N_2$) atoms in state $\ket{F=1}$ ($\ket{F=2}$).  Each pair of images represents the two extremes in the observed population ratio, $N_1/(N_1+N_2)$ (open circles in (c) and (d)). Population ratio variations between trials reflect interferometer phase variations caused by vibration of the retroreflection mirror.  Also shown in (a) and (b) are the atom densities integrated horizontally for the two images (black and red curves), with the shaded regions used to determine the port atom numbers, $N_i$. The lower port has been optically pushed, resulting in a hotter cloud with fewer peak counts.  Both ports are heated by a $5 \, \milli \seconds$ imaging pulse. This heating is most evident for $3~\nano \Kelvin$ clouds.}
\label{Fig:Interferometers}
\end{figure}

Figures \ref{Fig:Interferometers}a and \ref{Fig:Interferometers}b show images of both output ports for a rotation-compensated interferometer using two atom source temperatures. The interferometer in Fig.~\ref{Fig:Interferometers}a ($3~\nano\Kelvin$) has an integrated interferometer contrast of $80 \%$ while that in Fig.~\ref{Fig:Interferometers}b (50 nK) shows a contrast of $48 \%$ \footnote{Integrated contrast is calculated by summing image counts inside regions of interest around each output port and then forming the normalized population ratios $r_i$ for a set. The contrast of the set is $c=[\text{Max}(r_i)-\text{Min}(r_i)]/[\text{Max}(r_i)+\text{Min}(r_i)]$}. The contrast is reduced for the hotter source because of Rabi pulse area inhomogeneities due to larger horizontal cloud diameter (with respect to the spatially nonuniform laser beam intensity) and larger Doppler width.

With PSI, we maintain spatial fringe contrast even in the presence of large net rotation rates (Fig.~\ref{Fig:Contrast}). By comparison, the conventional integrated contrast for the same data decays rapidly with increasing rotation rate because a spatial average over the fringe pattern washes out the interference.  The reduction in the PSI fringe contrast at higher rotation rates is not fundamental, but results from heating during imaging and imperfect alignment between the applied rotation $\v{\delta\Omega}$ and the camera line-of-sight.

To compute spatial fringe contrast in Fig.~\ref{Fig:Contrast}, we divide the fitted amplitude of the population fringes by the fitted amplitude of the underlying cloud \cite{supplementalMaterial}. While fringes are visible on each raw image, we use Principal Component Analysis (PCA) as a filter to isolate the population fringe from the cloud shape in a model-independent way for more robust fits \cite{Segal2010}. The fitted fringe frequency provides the magnitude of the phase gradient.

\begin{figure}
\centering
\includegraphics[width=\columnwidth]{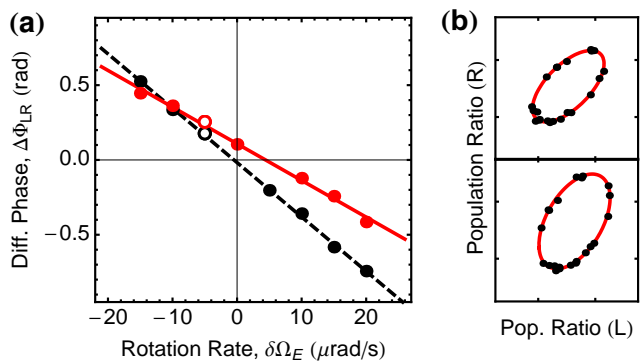}
\caption{(a) PSI dual-axis gyroscope.  We extract the differential phase $\Delta\Phi_{LR}$ between the left and right sides of the ensemble as a function of the rotation rate $\delta\Omega_{E}$, as measured on cameras CCD1 (black, dashed) and CCD2 (red, solid). (b) Sample ellipses emerging from the right-versus-left population ratios of CCD2 (upper) and CCD1 (lower), corresponding to the open circles of part (a).}
\label{Fig:GyroData}
\end{figure}

We also measure the rotation rate of the Earth.  After coarsely compensating for the Earth's rotation with the tip-tilt stage, we tune the applied rate by adding a small rotation $\delta\Omega_E\equiv\OmegaRCS-\OmegaEarth$ along the nominal direction of true North ($\phiRCS\approx\phiEarth + \pi$). We observe the resulting phase gradient simultaneously on CCD1 and CCD2.  The magnitude of the observed phase gradient depends on the projection of the net rotation rate onto each camera (see Fig.~\ref{Fig:Apparatus}c). To detect small phase gradients that generate less than $2\pi$ radians of phase across the ensemble, we extract the differential phase $\Delta\Phi_{LR}$ by splitting each image about a vertical line and analyzing the left and right halves as independent interferometers.

Figure \ref{Fig:GyroData}a shows $\Delta\Phi_{LR}$ as a function of $\delta\Omega_E$ as observed on CCD1 and CCD2. Each measurement is the result of 20 interferometer cycles.  We parametrically plot the population ratio of the left half versus the right (e.g., Fig. \ref{Fig:GyroData}b) and extract the differential phase and contrast using an ellipse fitting procedure \cite{Foster2002}. Occasional trials ($< 5\%$) that display no interference appear at the center of the ellipses and are rejected. The horizontal intercept of a linear fit to this data provides a measurement of Earth's rotation rate with a precision of $200~\nano\radians/\seconds$.

The difference in the intercepts observed by the two cameras indicates that the rotation compensation direction $\phiRCS$ is slightly misaligned from true North $\phiEarth$ such that $\Delta\phi\equiv\phiRCS-\left(\phiEarth+\pi\right)\neq 0$. This results in a spurious rotation $\left(\Delta\phi\,\OmegaEarth \sin \phiEarth \right)\hat{x}$ that imprints a phase gradient visible on CCD2 (see Table \ref{Tab:phases}, term 2) independent of $\delta\Omega_E$. Likewise, a spurious rotation $\left(-\Delta\phi\,\OmegaEarth \cos \phiEarth \right)\hat{y}$ imprints a phase gradient visible on CCD1. The slopes for the two cameras in Fig.~\ref{Fig:GyroData} are different because of unequal projection of $\v{\OmegaEarth}$ and small differences in the projected widths of the ensemble.

Although the mean interferometer phase is dominated by seismic noise contributions at long $T$, we can infer an acceleration sensitivity using the observed differential phase noise between different parts of the imaged cloud.  We divide the output port images using a checkerboard grid and study the differential phase between the combined even and combined odd grid squares. Varying the grid size $s$ in this analysis reveals correlated phase noise at different spatial scales \footnote{To ensure that results are independent of the initial grid registration, we compute two grid alignment quadratures (analogous to sine and cosine) for each dimension by offsetting the grid by $s/2$ in each direction. We then average over alignment using the root mean square of these four results.}. Analyzing $280$ trials with $\v{\OmegaRCS}\approx-\v{\OmegaEarth}$, we find the differential even-odd phase noise is $2.0~\text{mrad}$ per shot for grid sizes below $s=3~\text{mm}$. Combined with the acceleration response (Table \ref{Tab:phases}, term 1), this implies an acceleration sensitivity of $6.7 \times 10^{-12} g$ in one shot \footnote{The sensitivity is $\delta a/g=\delta\phi/\keff g T^2$, where $\delta\phi=(2.0~\milli\radians)/\sqrt{2}$ is the absolute phase noise combining all the atoms from both the even and odd grid squares.}, an improvement of more than two orders of magnitude over previous limits \cite{Muller2008}.  By comparison, the atom shot-noise limit for the $4\times 10^{6}$ atoms used in this interferometer at $50\%$ contrast is $\sim4\times 10^{-12}g$ in one shot. Note that this grid analysis rejects low spatial frequency variations of the phase across the cloud that originate, for example, from fluctuations in initial kinematics. The results are applicable to measurements where these effects are expected to be common, such as for overlapped ensembles of two species of atoms in an equivalence principle test.

PSI does not require a 10-meter apparatus. A dual-axis gyroscope with shot-noise-limited rotation noise of $100~\mu\text{deg}/\sqrt{\text{hour}}$ can be realized with $10^6$ atoms prepared at $3~\text{mK}$ in an interferometer with $T = 10 \, \milli \seconds$ and $4 \hbar k$ atom optics cycling at $25 \, \Hz$ (with atom recapture).

PSI can measure the interferometer beam optical wavefront in situ. This is desirable in precision atom interferometry applications, including gravitational wave detection \cite{Hogan2011}. Each atom in an expanding ensemble samples the laser phase at three locations, thereby measuring wavefront aberrations. Term 6 of Table~\ref{Tab:phases} models the interferometer response to a parabolic wavefront curvature of the form $k \alpha \left(x^2+y^2\right)/2$. Our measured phase noise implies a wavefront sensitivity of $\alpha\sim\frac{\lambda}{500}/ \centi\meters^2$ in one shot.

Finally, PSI allows measurement of multiple components of the gravitational gradient tensor (Table~\ref{Tab:phases}, term 5). The sensitivity we report is also sufficient to observe the gravity curvature induced phase shift (Table \ref{Tab:phases}, term 4) \cite{Audretsch1994}.
Such sensitivity enables precision tests of the equivalence principle and general relativity \cite{Hogan2009, Dimopoulos2008b}.

\begin{acknowledgments}
The authors would like to acknowledge the valuable contributions of Tim Kovachy, Sheng-wey Chiow, Jan Rudolph and Philippe Bouyer. AS acknowledges support from the NSF GRFP.  SMD acknowledges support from the Gerald J. Lieberman Fellowship. This work was supported in part by NASA GSFC Grant No. NNX11AM31A.
\end{acknowledgments}

\bibliographystyle{apsrev4-1}
\bibliography{EPKasevich-PSI}

\end{document}